\documentclass[12pt,preprint]{aastex}
\slugcomment{revised version, 2003.11.12}

\shorttitle{DUST TORUS IN NGC 4151}
\shortauthors{Minezaki et al.}

\begin{document}

\title{INNER SIZE OF A DUST TORUS
IN THE SEYFERT 1 GALAXY NGC 4151}

\author{Takeo Minezaki\altaffilmark{1,2},
    Yuzuru Yoshii\altaffilmark{1,3}, Yukiyasu Kobayashi\altaffilmark{4},
    Keigo Enya\altaffilmark{5},
    Masahiro Suganuma\altaffilmark{4,6}, Hiroyuki Tomita\altaffilmark{4,6}
    Tsutomu Aoki\altaffilmark{7},
    and
    Bruce A. Peterson\altaffilmark{8}}

\altaffiltext{1}{Institute of Astronomy, School of Science, University of Tokyo,
    2-21-1 Osawa, Mitaka, Tokyo 181-0015, Japan}
\altaffiltext{2}{e-mail: minezaki@mtk.ioa.s.u-tokyo.ac.jp}
\altaffiltext{3}{Research Center for the Early Universe,
    School of Science, University of Tokyo,
    7-3-1 Hongo, Bunkyo-ku, Tokyo 113-0033, Japan}
\altaffiltext{4}{National Astronomical Observatory,
    2-21-1 Osawa, Mitaka, Tokyo 181-8588, Japan}
\altaffiltext{5}{Institute of Space and Astronautical Science,
    Japan Aerospace Exploration Agency,
    3-1-1, Yoshinodai, Sagamihara, Kanagawa, 229-8510, Japan}
\altaffiltext{6}{Department of Astronomy, School of Science, University of Tokyo,
    7-3-1 Hongo, Bunkyo-ku, Tokyo 113-0013, Japan}
\altaffiltext{7}{Kiso Observatory,
    Institute of Astronomy, School of Science, University of Tokyo,
    10762-30 Mitake, Kiso, Nagano 397-0101, Japan}
\altaffiltext{8}{Mount Stromlo Observatory,
    Research School of Astronomy and Astrophysics,
    The Australian National University,
    Weston Creek P. O., A.C.T. 2611, Australia}

\begin{abstract}
   The most intense monitoring observations yet made
 were carried out on the Seyfert 1 galaxy NGC 4151
 in the optical and near-infrared wave-bands. 
 A lag from the optical light curve
 to the near-infrared light curve was measured.
 The lag-time between the $V$ and $K$ light curves
 at the flux minimum in 2001 was precisely
 $48^{+2}_{-3}$ days, as determined by a cross-correlation analysis.
 The correlation between the optical luminosity of an 
 active galactic nucleus (AGN)
 and the lag-time between the UV/optical and the near-infrared light curves
 is presented for NGC 4151 in combination with
 previous lag-time measurements 
 of NGC 4151 and other AGNs in the literature.
 This correlation is interpreted as
 thermal dust reverberation in an AGN, where
 the near-infrared emission from an AGN is expected to be
 the thermal re-radiation from hot dust surrounding the central engine
 at a radius
 where the temperature equals to that of the dust sublimation temperature. 
 We find that
 the inner radius of the dust torus in NGC 4151 is $\sim 0.04$ pc
 corresponding to the measured lag-time, well outside the 
 broad line region (BLR) determined by other reverberation studies
 of the emission lines.
\end{abstract}

\keywords{galaxies: Seyfert
 --- galaxies: active
 --- infrared: galaxies
 --- dust, extinction
 --- galaxies: individual(NGC 4151)}

\section{INTRODUCTION}
    One attractive idea of unifying active galactic nuclei (AGNs)
 of types Seyfert 1 and 2,
 depending on whether strong, broad emission lines are visible or not,
 hinges upon the existence of a dust torus
 that surrounds the broad-line region (BLR) \citep{Ant93}.
    A direct view of the BLR through the inner hole of the dust torus
 seen face-on yields a Seyfert 1 spectrum,
 whereas an obscuration of the BLR by the dust torus
 seen edge-on results in a Seyfert 2 spectrum. 
    Although this unified model has been supported in many ways already,
 the strongest justification would come from 
 confirming the existence of the dust torus surrounding the BLR.
    Since modern technology has not yet reached a point
 where one might spatially resolve the innermost structure of AGNs,
 the only method available to uncover it is a lag analysis in light curves,
 known as reverberation mapping.

    The near-infrared continuum ($\lambda \ga 2\ \mu$m) emission 
 of Seyfert galaxies is considered to
 be dominated by thermal radiation from hot dust
 surrounding the central engine,
 based on the decomposition of spectral energy distribution (SED)
 of AGNs (e.g. Kobayashi et al. 1993).
    The dust closest to the central engine has higher temperatures,
 but cannot survive inside some critical radius
 at which it begins to sublimate. 
    Accordingly, the distribution of dust should
 be torus-like, with a hole at the critical radius,
 and the dust at the innermost point of the torus
 has the maximum temperature equal to the sublimation temperature.
 This dust produces thermal emission which has a peak luminosity
 at the corresponding near-infrared wavelength.

    Clear evidence that the hot dust is located some distance
 from the center of the central engine in AGNs
 is the existence of a lag time between the UV or optical light curve
 and the near-infrared light curve of an AGN.
    It is interpreted in terms of light travel time from the central engine,
 which almost simultaneously emits UV and optical radiation
 to the surrounding hot dust,
 which reprocesses the radiation from the central engine
 to thermal near-infrared radiation.
     Previous studies reported such lags of $10-1000$ days
 for at most ten AGNs
 \citep{CWG89,glass92,sitko93,okn93,nelson96,okn99,bari92,nelson01,okn01},
 however, the number of targets and the measurement errors were somewhat limited.

     To make it possible to carry out
 much more intense monitoring observations of many AGNs
 in optical and the near-infrared
 the MAGNUM
 (\underline{M}ulticolor \underline{A}ctive \underline{G}alactic
  \underline{Nu}clei \underline{M}onitoring)
 project installed a 2-meter telescope
 at the Haleakala Observatories in Hawaii \citep{kob98a,yoshii02,yoshii03},
 and started its operation in early 2001.

     In this {\it Letter}, as an early result of the MAGNUM project, 
 we present the $V(0.55\ \mu\rm{m})$ and $K(2.2\ \mu\rm{m})$
 light curves of a bright Seyfert 1 galaxy NGC 4151
 for the period including the epoch of 
 flux minimum in 2001,
 and report a measurement of lag-time 
 between them based on a cross-correlation analysis.
 Due to the high frequency and high signal-to-noise ratio of our monitoring 
 observations, the measured lag is the most precise ever made,
 not only for NGC 4151 but also for any other AGNs.
 Then it was compared with the previously measured lag-times
 of other AGNs in literature,
 and interpreted by thermal dust reverberation in NGC 4151.
 The lag-time between the $V$ and $K$ light curves
 corresponds to the light travel-time
 between the central engine and the inner radius of the dust torus,
 and with this result, we argue in favor of the unified model that
 NGC 4151 has a dust torus whose innermost radius is well outside the BLR.

\section{OBSERVATION}

    Monitoring observations of NGC 4151 started January, 2001
 in the $V$ band and the $K$ band,
 using the multicolor imaging photometer (MIP)
 mounted on the MAGNUM telescope \citep{kob98a,kob98b}.
    The field of view of the MIP is $1.5\times 1.5$ arcmin${}^{2}$,
 and it enables simultaneous imaging in optical and near-infrared
 by splitting the incident beam into two different detectors,
 a SITe CCD ($1024\times 1024$ pixels, $0.277$ arcsec/pixel)
 and an SBRC InSb array ($256\times 256$ pixels, $0.346$ arcsec/pixel).

    We obtained photometric data for NGC 4151 between January and December 2001,
 consisting of $44$ nights in the $V$ band and $49$ nights in the $K$ band.
 With the exception of the three months during solar conjunction,
 from mid-August to the beginning of November,
 the average interval of the observations was less than a week. 
 This monitoring has provided
 the most frequent simultaneous optical and near-infrared 
 sampling of NGC 4151 to date.

    The telescope was pointed at NGC 4151 and two reference stars
 with ($\Delta \alpha,\Delta \delta) = (16.1\arcmin,5.7\arcmin$)
 and ($-11.3\arcmin,-9.3\arcmin$) alternately, and then it was dithered. 
 In this manner, high signal-to-noise ratios were achieved
 with total on-source integration time of a few minutes
 for each observation.
 The typical FWHMs of the point spread function (PSF)
 during the observation were
 1.5 arcsec in the $V$ band and 1.1 arcsec in the $K$ band, respectively.

     The images were reduced using IRAF
\footnote{
 IRAF is distributed by the National Optical Astronomy Observatories,
 which are operated by the Association of Universities for Research
 in Astronomy, Inc., under cooperative agreement with the National
 Science Foundation.
}
.
     We followed standard procedures for image reduction,
 with small corrections for the non-linear response of the detectors.
 For the infrared image reduction,
 a sky image assembled from dithered images pointed at the reference stars
 was subtracted prior to flat-fielding.
 Then, flat-field corrections were applied using dome flats for the optical images,
 or an illumination-corrected dome flats for the near-infrared images.

     The nuclear flux of NGC 4151 was measured relative to the reference stars,
 then the fluxes of these stars were calibrated.
     Aperture photometry within $\phi=8.3$ arcsec was applied to all images,
 then the fluxes of the NGC 4151 nucleus and the reference stars
 were compared for each dithering set.
 Finally, the nuclear flux relative to the reference stars
 was estimated from the average
 of the relative fluxes for all of the dithering sets.
 The photometric errors for data taken each night were
 usually better than $\sigma = 0.01$ mag.
 The flux ratio between the reference stars was also monitored;
 it was confirmed that they were not variable stars.

     The fluxes of the reference stars were calibrated
 with respect to photometric standard stars taken from \citet{landolt92} and \citet{hunt98},
 and a small correction for the filter color term in the $V$ band was applied.
 The flux calibration errors were
 less than 0.01 mag in $V$ and about 0.01 mag in $K$.
 The flux of the host galaxy within the aperture was estimated 
 by fitting the galaxy profile using the GALFIT \citep{peng02}.
 The flux of the host galaxy in the $V$ band was
 slightly dependent on the seeing FWHM,
 and was corrected for.
 The flux of the host galaxy was subtracted from the aperture flux;
 the errors arising from this subtraction were about 0.1 mag.
 We note that the subtraction provides only a flux offset,
 which does not affect the cross-correlation function analysis below.

\section{RESULTS}
\begin{figure}
\plotone{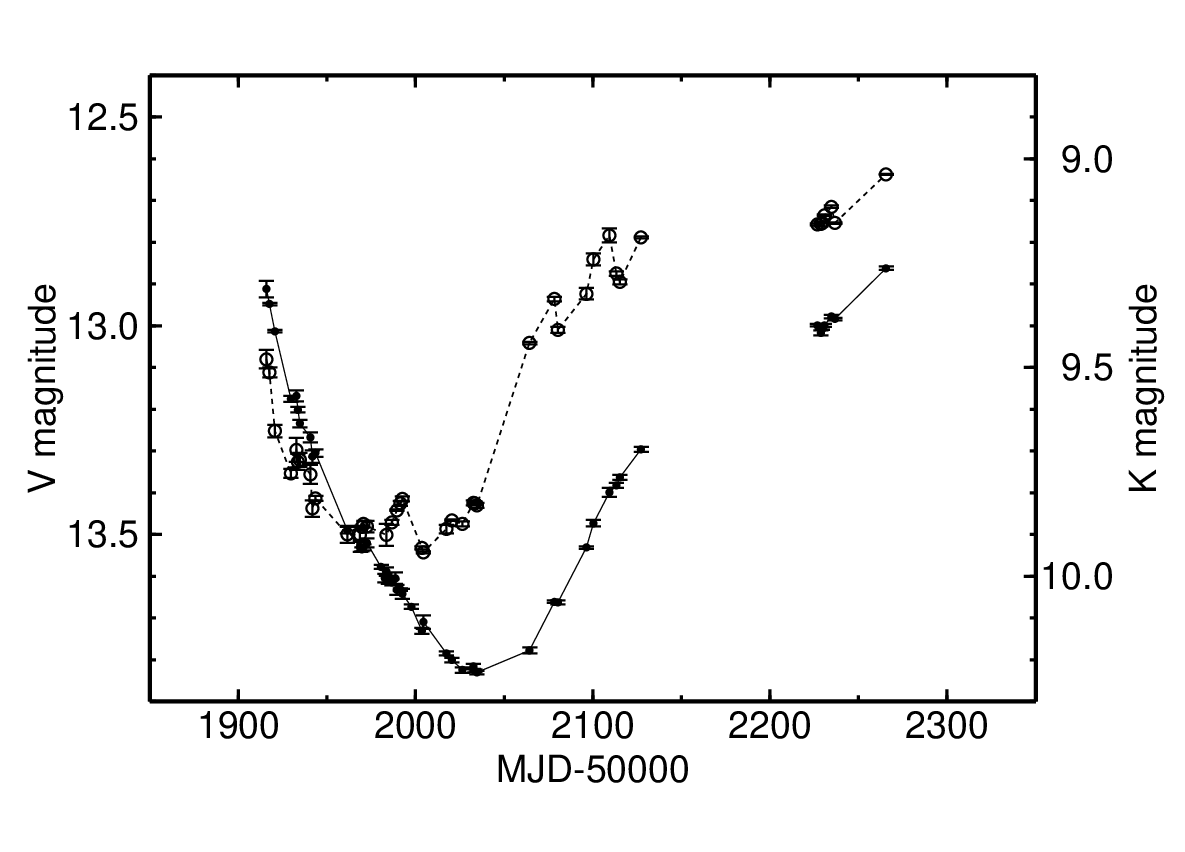}
\caption{The $V$ (open circles connected with dashed lines)
 and the $K$ (closed circles connected with solid lines)
 light curves of NGC 4151 nucleus in 2001.
 The flux from the host galaxy is subtracted.
 The flux minimum of the $K$ light curve
 is clearly delayed behind that of the $V$ light curve.
 The monitoring observation was interrupted due to the solar conjunction
 at ${\rm MJD}= 52130\sim 52220$.
 \label{fig1}}
\end{figure}

      Figure \ref{fig1} shows the $V$ and $K$ light curves
 of NGC 4151 nucleus in 2001. 
 The $V$ flux minimum was observed at epoch $\rm{MJD}\approx 51980$;
 the $K$ flux minimum was clearly delayed behind the $V$ flux minimum by about $50$ days.
 While the $V$ light curve exhibited some variability
 on short time-scales comparable to
 the typical monitoring interval of several days,
 the $K$ light curve changed smoothly for the whole
 period of the monitoring observations.
 This indicates that the $K$ band emitting region is
 extended more than a scale of several light-days.

     In order to obtain a more quantitative estimate of the lag-time
 between the $V$ and $K$ light curves,
 a cross-correlation analysis was applied.
 A cross-correlation function (CCF) of irregularly sampled photometric 
 data was computed based on the linear interpolation method \citep{gp87,wp94},
 as modified below.
       The first method is a ``bidirectionally interpolated'' CCF.
 Each flux data point of one band, $f_1(t_i)$, at an epoch $t_i$,
 is paired with the linearly interpolated flux of the other band,
 $f_2(t_i+\tau)$, with a lag-time, $\tau$,
 then the CCF$(\tau)$ is computed from all data pairs
 $(f_V(t_i),f_K(t_i+\tau))$ and $(f_V(t_j-\tau),f_K(t_j))$,
 not the average of two CCFs each from
 $(f_V(t_i),f_K(t_i+\tau))$ and $(f_V(t_j-\tau),f_K(t_j))$
 as is often done.
       The second method is to obtain an ``equally sampled'' CCF.
 Both the $V$ and $K$ light curves are linearly interpolated to produce
 the flux data points every day,
 then the CCF$(\tau)$ is computed between
 those equally sampled light curves.
       The third method is the MCCF
 \citep[the MCCF parameter MAX was 10 days]{okn93}.

   A Monte-Carlo simulation was carried out to estimate
 the lag-time error using the CCF analysis.
 Many CCFs were calculated from artificial light curves,
 constructed by interpolating the observed light curves.
 In the artificial light curves, we incorporated
 not only the photometric errors at the observed data points,
 but also the short time-scale flux variation between the observed data points
 which serves to increase the scatter of the light curves
 at the interpolated data points.
 The flux variation was incorporated
 so that the structure function of the artificial light curves
 reproduced that of the observed light curve.
 We note that the error estimation here incorporated
 only the random errors caused by the photometric error and the flux variation,
 and there may be some possible systematic error.
 The details of the CCF computation and the simulation are described in \citet{sug03}.

\begin{figure}
\plotone{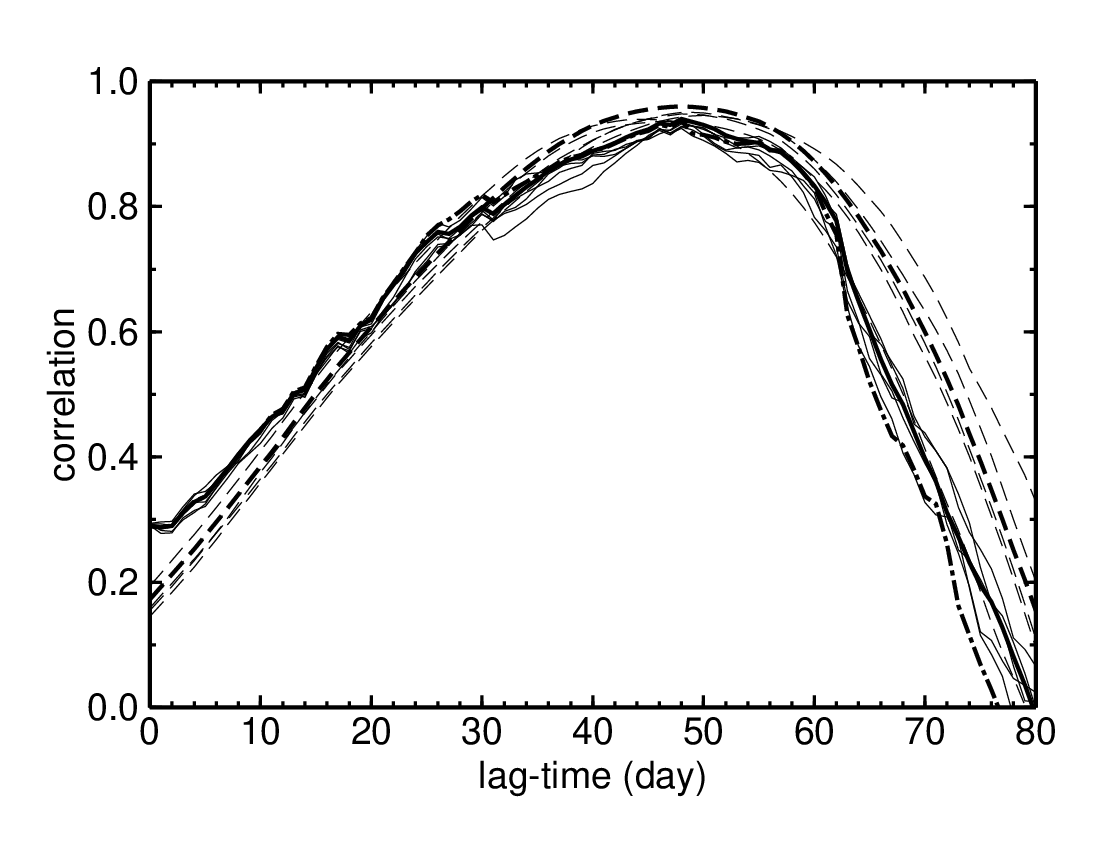}
\caption{The cross-correlation functions (CCFs) between the $V$ and $K$
 light curves of NGC 4151 at the flux minimum in 2001.
 The thick solid line is the ``bidirectionally interpolated'' CCF,
 the thick dashed line is the ``equally sampled'' CCF,
 and the thick dash-dotted line is the MCCF;
 all of them have a peak at a lag-time of $\Delta t=48$ days.
 The thin solid and dashed lines are the examples of CCFs calculated from
 the simulated light curves to estimate the lag-time errors.
 \label{fig2}}
\end{figure}

    Figure \ref{fig2} shows the CCFs
 between the $V$ and $K$ light curves of NGC 4151.
 Only the photometric data points
 (37 points in $V$ and 42 points in $K$)
 before the solar conjunction were used.
 All CCFs have a peak at the same lag-time of $\Delta t=48$ days.
 The average and the standard deviation of the lag-time at the CCF peak
 calculated from the simulation were
 $47.9$ and $1.1$ days for the ``bidirectionally interpolated'' CCF,
 $47.5$ and $2.4$ days for the ``equally sampled'' CCF,
 and $47.6$ and $0.9$ days for the MCCF.
     To be conservative, hereafter we use
 $\Delta t =48^{+2}_{-3}$ days as
 representative of the measured lag-time and the standard deviation error.
    The estimated lag-time of $\Delta t=48$ days indicate that
 the $K$ band emitting region was located
 at distance of 0.04 pc away from the central engine.

\section{DISCUSSION}
\begin{figure}
\plotone{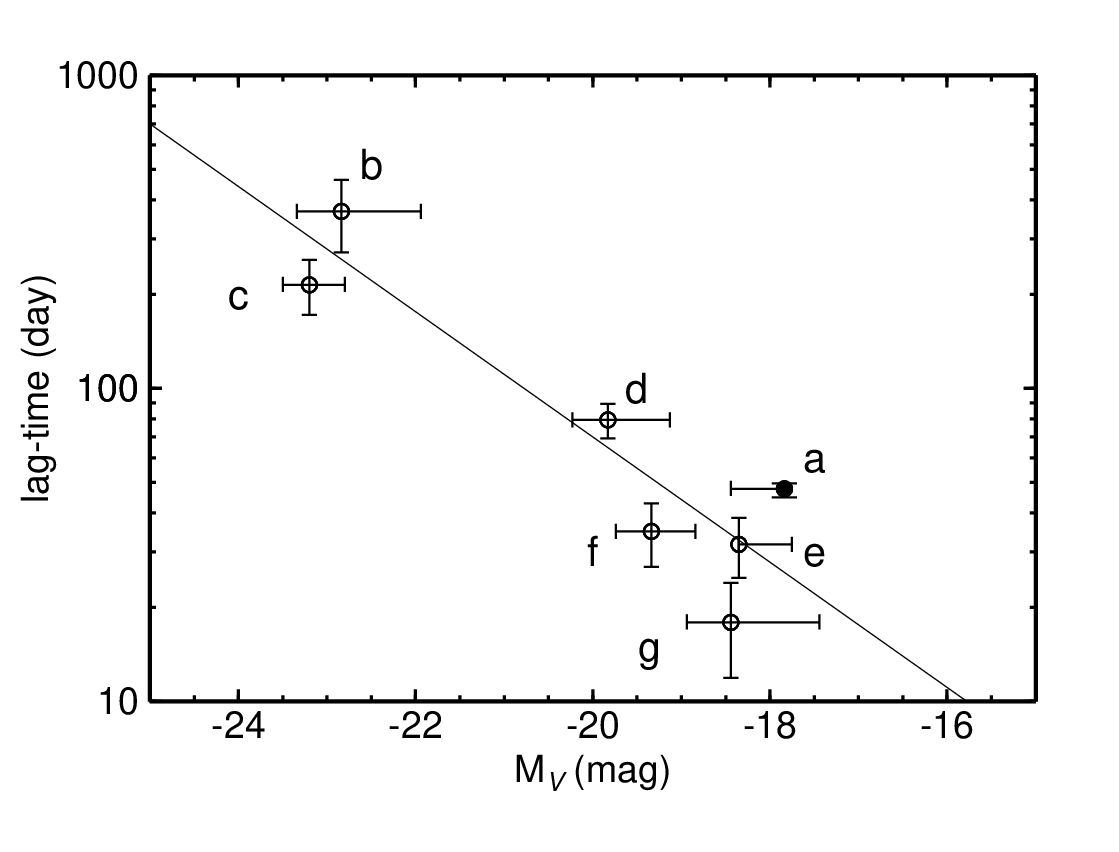}
\caption{The lag-times between the UV or optical
 and the $K$ light curves of AGNs
 against the absolute $V$ magnitudes.
 The closed circle (a) is this work (new data for NGC 4151);
 the open circles are taken from literature:
 (b) Fairall 9 \citep{CWG89}, (c) GQ Comae \citep{sitko93},
 (d) NGC 3783 \citep{glass92}, (e) Mrk 744 \citep{nelson96},
 (f) and (g) NGC 4151 \citep{okn99}.
 The inclined line was fitted assuming the expected slope from
 thermal dust reverberation, such that
 the lag-time is proportional to the square root of the luminosity.
 It should be noted that 
 the ``error bar'' in magnitude represents
 the observed range of variation during the observations.
 \label{fig3}}
\end{figure}

    Figure \ref{fig3} shows the lag-time of NGC 4151, together with
 those of other quasars and Seyfert 1 galaxies in the literature,
 plotted against their absolute $V$ magnitudes.
 The lag-times between the UV or optical light curve and the $K$ light curve
 were taken from the original references \citep{CWG89,glass92,sitko93,nelson96,okn99}.
 The lag-time data are restricted to those that were estimated
 by CCF analyses applied to the light curves.
 The absolute $V$ magnitudes were estimated from 
 their apparent magnitudes and redshifts in a range of 
 $cz=1000-50000$ km/s, assuming cosmological parameters of
 $(h_0,\Omega_0,\lambda_0)=(0.7,0.3,0.7)$ \citep{bennett03}.
     The flux of the host galaxy within the aperture was subtracted 
 by using the image decomposition technique, or the flux variation gradient
 \citep[\ ; this work]{bahcall97,kot93,winkler92,winkler97,glass92,nelson96}.
 Galactic extinction was corrected for \citep{schlegel98},
 but the intrinsic extinction in AGNs was ignored here for simplicity,
 because they were all Seyfert type 1 like AGNs.
 A K-correction was applied assuming a power-law spectrum 
 of $f_\nu\propto \nu^{-0.44}$ \citep{vanden01}.
 The observed lag-time was multiplied by a factor of $(1+z)^{-1}$
 to correct for the cosmological time dilation
\footnote{
 Assuming that the near-infrared emission at wavelength
 $\lambda $ is dominated by the thermal emission from
 heated dust of temperature $T$ which peaks at $\lambda $,
 the location of such dust at radius $r$ for given central
 luminosity scales as $r\propto T^{-2.8}$ \citep{bar87},
 leading to $\Delta t\propto \lambda^{2.8}$. Then, the
 lag-time correction for the cosmological wavelength
 dilation is expected to be a factor of $(1+z)^{2.8}$,
 opposite to that for the cosmological time dilation
 \citep{sitko93,okn01}.
 In the case of GQ Comae, having the largest redshift in
 the sample shown in Figure 3, this correction is equal
 to $(1+0.165)^{2.8}\approx 1.5$, placing the point (c)
 even closer to the inclined line.
}.

     It is clearly seen in Figure \ref{fig3} that the lag-time is correlated with 
 the absolute $V$ magnitude of the source, spanning over a range of 100 times 
 in luminosity and 10 times in lag-time,
 in the sense that a luminous AGN has a larger lag-time.
 A similar correlation has been previously noted by \citet{okn01},
 using the UV luminosity instead of $V$.
     The inclined line
 ($\log_{10} \Delta t = -2.15 - M_V/5.0$)
 shown in Figure \ref{fig3} is a fit to the data,
 with the expected slope of $\Delta t\propto L^{0.5}$
 from the reverberation of thermal emission
 of the hot dust surrounding the central engine.
     The inner radius of the dust torus is
 determined by the highest sublimation temperature of the dust,
 around $T\approx 1800$ K for graphite grains \citep{salpeter77} 
 or $T\approx 1500$ K for silicate grains \citep{huffman77}.
 The dust at the innermost limit of the torus produces
 strong thermal emission in the near-infrared.
     Since the sublimation radius is proportional to
 the square root of luminosity of the central engine \citep{bar87},
 the lag-time between the UV/optical and
 the $K$ light curves, which would corresponds to a light travel-time
 between the central engine and the inner radius of the dust torus,
 should be proportional to the square root of the luminosity of the central engine.
     Several studies have applied dust reverberation models to
 Fairall 9 \citep{bar92} and Mrk 744 \citep{nelson96},
 and estimated maximum temperatures for the circumnuclear dust
 to be comparable to the expected sublimation temperature.

     The good correlations that exist between the reported lag-times
 and the absolute $V$ magnitude along the expected slope
 from thermal dust reverberation,
 including our new result for NGC 4151,
 indicate that the $K$ band emission of NGC 4151 is likely to be dominated by
 the thermal radiation of hot dust at its innermost radius;
 the radius of the dust torus would be $\sim 0.04$ pc
 in order to correspond to the observed lag-time of $\Delta t = 48$ days.

     There are measurements of shorter lag-times for NGC 4151 in the past,
 such as $18\pm6$ days in $1969-1981$
 and $35\pm8$ days in $1993-1998$ by \citet{okn99}.
 Given a two-sigma level of their measurement error,
 then it may be possible to interpret that
 the recovering time-scale of the inner dust was so long,
 more than a few years,
 that the inner radius of the dust torus was not reduced
 after the most luminous state in $1993-1998$.
    However, the past data suffered from relatively large lag-time errors.
 In order to make it clear,
 it must be important to continue intense monitoring observation of NGC 4151
 to measure the lag-times at different epochs accurately,
 and in addition, monitoring the near-infrared color is also important
 because the dust temperature at the inner torus should change
 with flux variation
 when the inner radius of the dust torus is constant.

    Next, the measured lag-time for the dust torus is compared with
 similar estimates for the BLR,
 based on broad emission-line reverberation measurements.
 \citet{CWG89} observed not only the lag-times of near-infrared emission
 but also those of far-UV broad emission-lines,
 and concluded that the BLR lies inside the dust shell
 in the Seyfert 1 galaxy Fairall 9.
 Although we did not observe broad emission-lines at the same time,
 the measured lag-time of $48^{+2}_{-3}$ days for $K$ band emission was quite precise,
 and should correspond to the inner radius of the dust torus.
 Therefore, it is worthwhile to compare it with
 previous measurements of the lag-times of broad emission-lines.
     It is noted that the previous reports differ somewhat
 between the lines observed and the observations, 
 such as $\Delta t= 4\pm3 $ days for C{\small IV} and Mg{\small II} \citep{clavel90},
 $\Delta t=9\pm 2$ days for H{\small$\beta$} and H{\small $\alpha$} \citep{maoz91},
 $\Delta t=0-3$ days for H{\small $\beta$} and $\Delta t=0-2$ days for
 H{\small$\alpha$} \citep{kaspi96},
 $\Delta t=1.8-4.4$ days for C{\small IV} \citep{ulrich96},
 and $\Delta t=4\pm 2$ days for H{\small$\beta$} and H{\small $\alpha$} \citep{okn97}.

     It has been claimed that the spatial distribution of the BLR clouds 
 should be more extended than that estimated from the lag-times of
 the observed emission lines \citep{maoz91, ulrich96}.
 Even if this is the case, the inner radius of the
 dust torus in NGC 4151 is found to be well outside the BLR,
 therefore, the BLR in NGC 4151 can be observed directly
 without strong obscuration unless the dust torus is edge-on.
 Also the near-infrared emission dominated by thermal radiation
 from hot dust at the innermost region is observed
 without strong obscuration due to outer cold dust.
 These results are consistent with the unified model of Seyfert types of AGNs.

\section{CONCLUSION}
     An optical and near-infrared monitoring observation
 of the Seyfert 1 galaxy NGC 4151 was carried out,
 and a lag-time between the $V$ and $K$ light curves
 at the flux minimum in 2001
 was estimated at $\Delta t =48^{+2}_{-3}$ days
 by a cross-correlation analysis.
     We present that
 the lag-time between the UV or optical and the $K$ light curves
 is correlated with the absolute $V$ magnitude of AGN,
     and conclude that
 the $K$ band emission of NGC 4151
 is dominated by the thermal radiation of
 innermost hot dust whose temperature is comparable to dust sublimation,
 and that the inner radius of the dust torus of NGC 4151
 is $\sim 0.04$ pc, which is well outside the BLR.

     The ongoing intense multicolor monitoring observations of NGC 4151
 and many other AGNs in the MAGNUM project
 will not only study the detailed geometry of the dust tori
 and further examine the unified model,
 but also constrain
 the real-time change in size of the dust tori and the time-scales of
 dust formation and destruction under extreme circumstances in AGNs
 from measurements of $\Delta t$ at different epochs along the light curves
 and the near-infrared color with flux variation.

\acknowledgments

 We thank Mamoru Doi and Kentaro Motohara
 for helpful discussions and advices during the observations,
 and also thank colleagues at the Haleakala Observatories
 for their help for the facility maintenance.
 We are grateful to Timothy C. Beers for his useful comments.
 This research has been supported partly by
 the Grant-in-Aid of Scientific Research
 (10041110, 10304014, 12640233, 14047206, 14253001, 14540223)
 and COE Research (07CE2002) of
 the Ministry of Education, Science, Culture and Sports of Japan.

\clearpage 

\clearpage 

\begin{thebibliography}{}
\bibitem[Antonucci (1993)]{Ant93}Antonucci, R.
 1993, \araa, 31, 473
\bibitem[Bahcall et al. (1997)]{bahcall97}Bahcall, J. N.,
 Kirhakos, S., Saxe, D. H., \& Schneider, D. P.
 1997, \apj, 479, 642
\bibitem[Baribaud et al. (1992)]{bari92}Baribaud, T.,
 Alloin, D., Glass, I., Pelat, D.
 1992, \aap, 256, 375
\bibitem[Barvainis (1992)]{bar92}Barvainis, R.
 1992, \apj, 400, 502
\bibitem[Barvainis (1987)]{bar87}Barvainis, R.
 1987, \apj, 320, 537
\bibitem[Bennett et al. (2003)]{bennett03}Bennett, C. L. et al.
 2003, \apjs, 148, 97
\bibitem[Clavel, Wamsteker \& Glass (1989)]{CWG89}Clavel, J.,
 Wamsteker, W., \& Glass, I. S.
 1989, \apj, 337, 236
\bibitem[Clavel et al. (1990)]{clavel90}Clavel, J. et al.
 1990, \mnras, 246, 668 
\bibitem[Gaskell \& Peterson (1987)]{gp87}Gaskell, C. M.,
 \& Peterson, B. M.
 1987, \apjs, 65, 1
\bibitem[Glass (1992)]{glass92}Glass, I. S.
 1992 , \mnras, 256, 23P
\bibitem[Huffman (1977)]{huffman77}Huffman, D. R.
 1977, Adv. Phys., 26, 129
\bibitem[Hunt et al. (1998)]{hunt98}Hunt, L. K.,
 Mannucci, F., Testi, L., Migliorini, S., Stanga, R. M.,
 Baffa, C., Lisi, F., \& Vanzi, L.
 1998, \aj, 115, 2594
\bibitem[Kaspi et al. (1996)]{kaspi96}Kaspi, S. et al.
 1996, \apj, 470, 336
\bibitem[Kobayashi et al. (1993)]{kob93} Kobayashi, Y.,
 Sato, S., Yamashita, T., Shiba, H., \& Takami, H.
 1993, \apj, 404, 94
\bibitem[Kobayashi et al. (1998a)]{kob98a} Kobayashi, Y. et al.
 1998a, Proc. SPIE Vol. 3352, pp120-128, 
 Advanced Technology Optical/IR Telescopes VI, ed. L. M. Stepp
\bibitem[Kobayashi et al. (1998b)]{kob98b} Kobayashi, Y.,
 Yoshii, Y., Peterson, B. A., Minezaki, T.,
 Enya, K., Suganuma, M.,
 \& Yamamuro, T.
 1998b, Proc. SPIE Vol. 3354, pp769-776, 
 Infrared Astronomical Instrumentation, ed. A. M. Fowler
\bibitem[Kotilainen, Ward, \& Williger (1993)]{kot93}Kotilainen, J. K.,
 Ward, M. J., \& Williger, G. M.
 1993, \mnras, 263, 655
\bibitem[Landolt (1992)]{landolt92}Landolt, A. U.
 1992, \aj, 104, 340
\bibitem[Maoz et al. (1991)]{maoz91}Maoz, D. et al.
 1991, \apj, 367, 493
\bibitem[Nelson (1996)]{nelson96}Nelson, B. O.
 1996, \apjl, 465, L87
\bibitem[Nelson \& Malkan (2001)]{nelson01}Nelson, B. O.,
 \& Malkan, M. A.
 2001, ASP. Conf. Ser. Vol. 224, pp141-147,
 Probing the Physics of Active Galactic Nuclei
 by Multiwavelength Monitoring,
 ed. B. M. Peterson, R. S. Polidan, \& R. W. Pogge
\bibitem[Oknyanskij \& Horne (2001)]{okn01}Oknyanskij, V. L.,
 \& Horne, K.
 2001, ASP. Conf. Ser. Vol. 224, pp149-158,
 Probing the Physics of Active Galactic Nuclei
 by Multiwavelength Monitoring,
 ed. B. M. Peterson, R. S. Polidan, \& R. W. Pogge
\bibitem[Oknyanskij et al. (1999)]{okn99}Oknyanskij, V. L.,
 Lyuty, V. M., Taranova, O. G., \& Shenavrin, V. I.
 1999, Astronomy Letters, 25, 483
\bibitem[Oknyanskij \& van Groningen (1997)]{okn97}Oknyanskij, V. L.,
 \& van Groningen, E.
 1997, Astron. and Astrophys. Transactions, 14, 105
\bibitem[Oknyanskij (1993)]{okn93}Oknyanskij, V. L.
 1993, Astronomy Letters, 19, 416
\bibitem[Peng et al. (2002)]{peng02} Peng, C. Y.,
 Ho, L. C., Impey, C. D., \& Rix, H.-W.
 2002, \aj, 124, 266
\bibitem[Salpeter (1977)]{salpeter77} Salpeter, E. E.
 1977, \araa, 15, 267
\bibitem[Schlegel, Finkbeiner, \& Davis (1998)]{schlegel98}Schlegel, D. J.,
 Finkbeiner, D. P., \& Davis, M.
 1998 \apj, 500, 525
\bibitem[Sitko et al. (1993)]{sitko93} Sitko, M. L.,
 Sitko, A. K., Siemiginowska, A., \& Szczerba, R.
 1993, \apj, 409, 139
\bibitem[Suganuma et al. (2003)]{sug03} Suganuma, M. et al. in preparation
\bibitem[Ulrich \& Horne (1996)]{ulrich96} Ulrich, M.-H.,
 \& Horne, K.
 1996, \mnras, 283, 748
\bibitem[Vanden Berk et al. (2001)]{vanden01}
 Vanden Berk, D. E. et al.
 2001 \aj, 122, 549
\bibitem[White \& Peterson (1994)]{wp94} White, R. J.
 \& Peterson, B. M.
 1994, \pasp, 106, 879
\bibitem[Winkler et al. (1992)]{winkler92}Winkler, H.,
 Glass, I. S., van Wyk, F., Marang, F., Jones, J. H. S.,
 Buckley, D. A. H., \& Sekiguchi, K.
 1992, \mnras, 257, 659
\bibitem[Winkler (1997)]{winkler97}Winkler, H.
 1997, \mnras, 292, 273
\bibitem[Yoshii (2002)]{yoshii02}Yoshii, Y.
 2002, in New Trends in Theoretical and Observational Cosmology,
 ed. K. Sato and T. Shiromizu (Tokyo: Universal Academy Press), 235
\bibitem[Yoshii, Kobayashi \& Minezaki (2003)]{yoshii03}Yoshii, Y.,
 Kobayashi, Y., \& Minezaki, T.
 2003, AAS, 202, 3803
\end{thebibliography}
\end{document}